\documentclass[twocolumn]{aastex631}

\usepackage{mathtools}

\shorttitle{NSdelta}
\shortauthors{LI et al.}

\graphicspath{{./}{figures/}}

\begin{document}

\title{Effects of isoscalar- and isovector-scalar meson mixing on neutron star structure}

\correspondingauthor{Lie-Wen Chen}
\email{lwchen@sjtu.edu.cn}

\author{Fan Li}
\affiliation{School of Physics and Astronomy, Shanghai Key Laboratory for
Particle Physics and Cosmology, and Key Laboratory for Particle Astrophysics and Cosmology (MOE),
Shanghai Jiao Tong University, Shanghai 200240, China}

\author{Bao-Jun Cai}
\affiliation{Quantum Machine Learning Laboratory, Shadow Creator Inc., Shanghai 201208, China}

\author{Ying Zhou}
\affiliation{Quantum Machine Learning Laboratory, Shadow Creator Inc., Shanghai 201208, China}

\author[0000-0002-9188-5558]{Wei-Zhou Jiang}
\affiliation{School of Physics, Southeast University, Nanjing 211189, China}

\author[0000-0002-7444-0629]{Lie-Wen Chen}
\affiliation{School of Physics and Astronomy, Shanghai Key Laboratory for
Particle Physics and Cosmology, and Key Laboratory for Particle Astrophysics and Cosmology (MOE),
Shanghai Jiao Tong University, Shanghai 200240, China}

\begin{abstract}
Based on the accurately calibrated interaction FSUGold, we show that including isovector scalar $\delta$ meson and its coupling to isoscalar scalar $\sigma$ meson in the relativistic mean field~(RMF) model can soften the symmetry energy $E_{\rm{sym}}(n)$ at intermediate densities while stiffen the $E_{\rm{sym}}(n)$ at high densities. We find this new RMF model can be simultaneously compatible with
(1) the constraints on the equation of state of symmetric nuclear matter at suprasaturation densities from flow data in heavy-ion collisions,
(2) the neutron skin thickness of $^{208}$Pb from the PREX-II experiment,
(3) the largest mass of neutron star (NS) reported so far from PSR J0740+6620,
(4) the limit of $\Lambda_{1.4}\leq580$ for the dimensionless tidal deformability of the canonical 1.4$M_{\odot}$ NS from the gravitational wave signal GW170817,
(5) the mass-radius relation of PSR J0030+0451 and PSR J0740+6620 measured by NICER,
and thus remove the tension between PREX-II and GW170817 observed in the conventional RMF model.
\end{abstract}

\section{Introduction}
A model-independent extraction of the neutron skin thickness of $^{208}$Pb is reported recently by the updated Lead Radius EXperiment~(PREX-II)~\citep{2021PhRvL.126q2502A}:
\begin{equation}
\Delta r_{\rm{np}} \equiv r_n-r_p=0.283 \pm 0.071 ~\rm{fm}^{-3},
\end{equation}
where $r_n~({r_p})$ is the rms radii of the neutron (proton) distribution in the nucleus.
This value means a remarkably thick $\Delta r_{\rm{np}}$ for $^{208}$Pb compared to other results, and suggests a very stiff nuclear symmetry energy $E_{\rm{sym}}(n)$ around nuclear saturation
density $n_0 \sim 0.16$~fm$^{-3}$. The $E_{\rm{sym}}(n)$ is an essential but poorly-known quantity that encodes the isospin dependence of nuclear
matter equation of state (EOS) and plays a key role in nuclear physics and astrophysics~\citep{2005PhR...410..335B,2005PhR...411..325S,2008PhR...464..113L,2015ARNPS..65..303G,2016ARA&A..54..401O,2016PrPNP..91..203B,2019EPJA...55...39Z}. Indeed, it has been established that the symmetry energy density slope $L(n)$ around $n_0$ exhibits a strong positive linear correlation with the $\Delta r_{\rm{np}}$ of finite nuclei~\citep{2000PhRvL...85...5296R,2002NuPhA.706...85F,2005PhRvC..72f4309C,2009PhRvL.102l2502C,2010PhRvC..82b4321C,2011PhRvL.106y2501R}. In particular, there is an even stronger linear correlation between the $\Delta r_{\rm{np}}$ and the $L(n_c)$ at a subsaturation cross density $n_c = 0.11/0.16n_0 \approx 2/3n_0$ (roughly corresponding to the average density of finite nuclei), and the $\Delta r_{\rm{np}}$ of heavy nuclei can be uniquely determined by $L(n_c)$~\citep{2013PhLB..726..234Z}. The large $\Delta r_{\rm{np}}$ of $^{208}$Pb from PREX-II requires $L(n_c) \gtrsim 52$~MeV~\citep{2021arXiv210205267Y} or $L(2/3n_0) \gtrsim 49$~MeV~\citep{2021PhRvL.126q2503R}, suggesting a quite stiff $E_{\rm{sym}}(n)$ at subsaturation density.

The stiff $E_{\rm{sym}}(n)$ usually results in a large value of neutron star~(NS) radius and tidal deformability. Using the $\Delta r_{\rm{np}} =0.283 \pm 0.071 ~\rm{fm}^{-3}$ of $^{208}$Pb from PREX-II, Reed {\it et al.}~\citep{2021PhRvL.126q2503R} have recently obtained a constraint of $642\leqslant \Lambda_{1.4} \leqslant 955$ for the dimensionless tidal deformability of 1.4$M_{\odot}$ NS within the non-linear relativistic mean field~(RMF) model.
However, a limit of $\Lambda_{1.4}$ $\leq 580$ has been extracted from the gravitational wave signal GW170817~\citep{2018PhRvL.121p1101A}, indicating a serious tension between the PREX-II experiment and GW170817 in the RMF model, although this tension has not been observed in the non-relativistic energy density functional with the extended Skyrme interactions including the momentum dependence of effective many-body forces~\citep{2021arXiv210205267Y}.
Also, we note that the tension between PREX-II and GW170817 can be ameliorated in Bayesian analysis with parameterized EOSs of nuclear matter and NS matter~\citep{2021ApJ...921...63B}.
On the other hand, in the terrestrial lab, the EOS of symmetric
nuclear matter~(SNM) at suprasaturation densities has been constrained
by collective flow data in heavy-ion collisions~(HIC)~\citep{2002Sci...298.1592D}, which strongly restricts the NS maximum mass $M_{\rm max}$
with $M_{\rm max}\sim 2.3M_{\odot}$~\citep{2019PhRvD..99l1301Z,2019ApJ...886...52Z,2014arXiv1402.4242C}.
In addition, the heaviest NS with mass $2.08^{+0.07}_{-0.07}M_{\odot}$ observed so far for PSR J0740+6620~\citep{2020NatAs...4...72C,2021ApJ...915L..12F} can put strong constraints on the dense matter EOS and rule out the supersoft high density symmetry energy~\citep{2019ApJ...886...52Z}.
Furthermore, the simultaneous mass-radius~(M-R) determinations from Neutron Star Interior Composition Explorer~(NICER) for PSR J0030+0451~\citep{2019ApJ...887L..21R,2019ApJ...887L..24M} with mass around $1.4M_{\odot}$ as well as for PSR J0740+6620~\citep{2021ApJ...918L..27R,2021ApJ...918L..28M} with mass around $2.0M_{\odot}$ have been obtained, which further put stringent constraints on the symmetry energy.
With so many constraints/data in the multimessenger era, it is of extreme importance to test the applicability of nuclear energy density functionals, including relativistic and non-relativistic models.

The RMF models are covariant effective theory and this relativistic covariant formulation has made great success during the last decades in understanding many nuclear phenomena (see, e.g., Refs.~\citep{1986ANP...16...1A,1997IJMPE...6..515S,2006PrPNP..57..470M}).
Based on the Walecka model~\citep{1974AnnalsP...83..491W} and its extensions~\citep{1977NPA...292..413X,1979PhLB...86..146S,1996NuPhA.606..508M}, the RMF models usually involve the Dirac nucleons, the isoscalar scalar meson $\sigma$, isovector scalar meson $\omega$ and isovector vector meson $\rho$. In addition, the $\sigma$ meson self-interactions were
used~\citep{1977NPA...292..413X} to reduce the nuclear incompressibility coefficient of symmetric nuclear matter while the $\omega$ meson self-interactions were introduced to efficiently tune the NS maximum mass without sacrificing the agreement with other well-reproduced
observables~\citep{1996NuPhA.606..508M}.
Furthermore, the crossing term $\rho$-$\omega$ is highly sensitive to the density dependence of the symmetry energy and is important for the properties of neutron-rich nuclei as well as the structure and cooling dynamics of
NSs~\citep{2001PhRvL..86.5647H,2001PhRvC..64f2802H,2003ApJ...593..463C}.

According to relativistic covariance in the effective theory,
the isovector scalar meson $\delta$ should also be included in the Lagrangian of the RMF model, and it may play an important role in understanding the properties of  isospin asymmetric matter at high densities~\citep{1997PhLB..399..191K,2001PhRvC..64c4314H,2002PhRvC..65d5201L,2007PhRvC..76e4316C}.
Very recently, the coupling of $\delta$ meson with $\sigma$ meson has been studied by Zabari {\it et al.}~\citep{2019PhRvC..99c5209Z,2019PhRvC.100a5808Z}, and it is found that the $\delta$-$\sigma$ mixing can soften the $E_{\rm{sym}}(n)$ at intermediate densities but stiffen the $E_{\rm{sym}}(n)$ at high densities. This interesting feature provides the possibility to decrease the $\Lambda_{1.4}$ to be compatible with the constraint of $\Lambda_{1.4}\leq580$~\citep{2018PhRvL.121p1101A} since the $\Lambda_{1.4}$ is essentially controlled by the density behaviors of $E_{\rm{sym}}(n)$ at intermediate densities around $2n_0$.
Nevertheless, we note that
the parameter sets used in Ref.~\citep{2019PhRvC..99c5209Z} violate the constraints on the EOS of SNM at suprasaturation densities from flow data in heavy-ion collisions~\citep{2002Sci...298.1592D}, predict a $\Lambda_{1.4}$ larger than $580$. Also in Ref.~\citep{2019PhRvC..99c5209Z}, the $\omega$ meson self-interactions and the crossing term $\rho$-$\omega$ are not considered. A similar study has been performed very recently in Ref.~\citep{2022arXiv220206468M}.

In this work, based on the accurately calibrated interaction FSUGold~\citep{2005PhRvL..95l2501T}, we include the $\delta$ meson and the crossing term $\delta$-$\sigma$ in the RMF model, and find that the $\delta$-$\sigma$ mixing can soften the $E_{\rm{sym}}(n)$ at intermediate densities while stiffen the $E_{\rm{sym}}(n)$ at high densities. Our results indicate that the new RMF model can remove the tension between the PREX-II and GW170817 in the conventional RMF model, and at the same time it is compatible with
the constraints on the EOS of SNM at suprasaturation densities from flow data in HIC, the largest NS mass reported so far from PSR J0740+6620, and the M-R relation of PSR J0030+0451 and PSR J0740+6620 measured by NICER.

\section{Formalism}
The EOS of isospin asymmetric matter, defined by its binding energy per nucleon, can be expressed as
\begin{equation}
E(n,\alpha) = E_0(n)+E_{\rm{sym}}(n)\alpha^2+O(\alpha^4)
\end{equation}
where $n=n_n+n_p$ is the baryon density with $n_n$~($n_p$) denoting the neutron (proton) density, $\alpha = (n_n-n_p)/(n_n+n_p)$ is the isospin asymmetry, $E_0(n) = E(n,\alpha=0)$ is the EOS of SNM, and the symmetry energy $E_{\rm{sym}}(n)$ is defined as
\begin{equation}
\label{Eq:Esym_def}
E_{\rm{sym}}(n)=\frac{1}{2}\frac{\partial^2E(n,\alpha)}{\partial\alpha^2}\mid_{\alpha=0}.
\end{equation}
At nuclear matter saturation density $n_0$, the $E_0(n)$ can be expanded in $\chi \equiv (n-n_0)/3n_0$ as
\begin{equation}
\label{E_0expend}
E_0(n) = E_0(n_0)+\frac{K_0}{2!}\chi^2+\frac{J_0}{3!}\chi^3+O(\chi^4)
\end{equation}
where $E_0(n_0)$ is the binding energy per nucleon in SNM at $n_0$, $K_0=9n^2_0\frac{\partial^2E_0(n)}{\partial n^2}\mid_{n=n_0}$ is the incompressibility coefficient, and $J_0=27n^3_0\frac{\partial^3E_0(n)}{\partial n^3}\mid_{n=n_0}$ is the skewness coefficient. Similarly, the $E_{\rm{sym}}(n)$ can be expanded at a reference density $n_r$ as
\begin{equation}
\label{E_sym-expend}
E_{\rm{sym}}(n) = E_{\rm{sym}}(n_r)+L(n_r)\chi_r+O(\chi_r^2)
\end{equation}
with $\chi_r=(n-n_r)/(3n_r)$, and the symmetry energy density slope $L(n_r)$ at $n_r$ can be obtained as
\begin{equation}
\label{Eq:L_def}
L(n_r)=3n_r\frac{\partial E_{\rm{sym}}(n)}{\partial n}\mid_{n=n_r}.
\end{equation}
The conventional $L \equiv L(n_0)$ is recovered by setting $n_r = n_0$.
In particular, the $L_c \equiv L(n_c)$ is obtained at a subsaturation cross density $n_c = 0.11/0.16n_0$, which essentially determines the neutron skin thickness of $^{208}$Pb~\citep{2013PhLB..726..234Z}.

Based on the accurately calibrated interaction FSUGold~\citep{2005PhRvL..95l2501T}, we consider here the isovector scalar $\delta$ meson and its coupling to the isoscalar scalar $\sigma$ meson in the RMF model. The Lagrangian density $\mathcal{L}$ of the nucleon system is given by
\begin{eqnarray}
\label{Eq:Lagrangian}
\mathcal{L}&=&\overline{\psi}\left(i \partial_{\mu} \gamma^{\mu}-m\right) \psi\notag\\
&+&g_{\sigma} \sigma \overline{\psi} \psi-g_{\omega} \omega_{\mu} \overline{\psi} \gamma^{\mu} \psi-g_{\rho} \vec{\rho}_{\mu} \overline{\psi} \gamma^{\mu} \vec{\tau} \psi+g_{\delta} \vec{\delta} \overline{\psi} \tau \psi\notag\\
&+&\frac{1}{2}\left(\partial_{\mu} \sigma \partial^{\mu} \sigma-m_{\sigma}^{2} \sigma^{2}\right)-\frac{1}{3} b_{\sigma} m\left(g_{\sigma} \sigma\right)^{3}-\frac{1}{4} c_{\sigma}\left(g_{\sigma} \sigma\right)^{4}\notag\\
&-&\frac{1}{4} \omega_{\mu \nu} \omega^{\mu \nu}+\frac{1}{2} m_{\omega}^{2} \omega_{\mu} \omega^{\mu}+\frac{1}{4}c_{\omega}\left(g_{\omega}^{2} \omega_{\mu} \omega^{\mu}\right)^2\notag\\
&-&\frac{1}{4} \rho_{\mu \nu} \rho^{\mu \nu}+\frac{1}{2} m_{\rho}^{2} \vec{\rho}_{\mu} \vec{\rho}^{\mu}+\frac{1}{2}\Lambda_V(g_{\rho}^{2}\vec{\rho}_{\mu} \vec{\rho}^{\mu})(g_{\omega}^{2} \omega_{\mu} \omega^{\mu})\notag\\
&+&\frac{1}{2}\left(\partial_{\mu} \vec{\delta} \partial^{\mu} \vec{\delta}-m_{\delta}^{2} \vec{\delta}^{2}\right)+\frac{1}{2} C_{\delta \sigma} (g_{\sigma}^{2}\sigma^{2}) (g_{\delta}^{2}\vec{\delta}^{2})
\end{eqnarray}
where $m$ is the nucleon mass, $m_\sigma$,$m_\omega$,$m_\rho$ and $m_\delta$ are the meson masses, and $g_\sigma$,$g_\omega$,$g_\rho$ and $g_\delta$ represent coupling constants for nucleons with the corresponding mesons. In addition, $b_\sigma$ and $c_\sigma$ are dimensionless for the self-interaction of the $\sigma$ meson, $c_\omega$ is for the self-interaction of the $\omega$ meson, $\Lambda_V$ is the coupling coefficient between $\rho$ meson and $\omega$ meson, and $C_{\delta \sigma}$ is the coupling coefficient between $\sigma$ meson and $\delta$ meson. We note that the Lagrangian density of the FSUGold interaction (without including the electromagnetic interactions) can be obtained by removing the terms associated with $\delta$ meson in Eq.~(\ref{Eq:Lagrangian}).

With the standard Euler-Lagrange equation
\begin{equation}
\partial_\mu(\frac{\partial\mathcal{L}}{\partial(\partial_\mu \phi)})-\frac{\partial\mathcal{L}}{\partial\phi}=0,
\end{equation}
one can obtain equations of motion for
the nucleon and meson fields from the Lagrangian density. The resulting Dirac equation for the nucleon field is
\begin{equation}
[\gamma^\mu(i\partial_\mu+\Sigma_{\mu}^{J})-(m+\Sigma_{s}^{J})]\Psi=0
\end{equation}
with the nucleon scalar self-energy and vector self-energy given, respectively, by
\begin{subequations}
	\begin{align}
	\Sigma_{s}^{J}=&-g_\sigma\sigma-g_\delta\vec{\delta}\cdot\vec{\tau}, \\
	\Sigma_{\mu}^{J}=&g_\omega\omega_\mu+g_\rho\vec{\rho}_\mu\cdot\vec{\tau}.
	\end{align}
\end{subequations}
The equations of motion for the meson fields can be obtained as
\begin{eqnarray}
(\partial_\mu \partial^\mu+m^2_\sigma)\sigma&=&g_\sigma[\overline{\Psi}\Psi-b_\sigma M(g_\sigma \sigma)^2-c_\sigma (g_\sigma \sigma)^3\notag\\
&&+C_{\delta\sigma}(g_\sigma \sigma)(g_\delta \vec{\delta})^2],\\
\partial_\mu \omega^{\mu \nu}+m^2_\omega \omega^\nu&=&g_\omega[\overline{\Psi}\gamma^\nu\Psi-c_\omega g_\omega^3\omega_\mu\omega^\mu\omega^\nu\notag\\
&&-\Lambda_V (g_\omega\omega^\nu) g_\rho^2\vec{\rho}_{\mu} \vec{\rho}^{\mu}],\\
\partial_\mu\vec{\rho}^{\mu \nu}+m^2_\rho\vec{\rho}^{\nu}&=&g_\rho[\overline{\Psi}\gamma^\nu\vec{\tau}\Psi-\Lambda_V (g_\rho \vec{\rho}^{\nu})g_\omega^2\omega_\mu\omega^\mu],\\
(\partial_\mu \partial^\mu+m^2_\delta)\vec{\delta}&=&g_\delta[ \overline{\Psi}\vec{\tau}\Psi+C_{\delta\sigma}(g_\delta\vec{\delta})(g_\sigma\sigma)^2].
\end{eqnarray}

For a static and homogenous infinite nuclear matter, all derivative terms drop out and the expectation values of spacelike components of vector fields vanish (only zero components $\vec{\rho}_0$ and $\omega_0$ survive) due to translational invariance and rotational
symmetry of the nuclear matter. In addition, only the third component of isovector fields ($\delta^{(3)}$ and $\rho^{(3)}$) needs to be considered because of the rotational invariance around the third axis in the isospin space. In the mean-field approximation, meson fields are replaced by their expectation values, i.e., $\sigma \rightarrow \overline{\sigma}$, $\omega_\mu \rightarrow \overline{\omega}_0$, $\rho_\mu \rightarrow \overline{\rho}^{(3)}_0$, and $\delta \rightarrow \overline{\delta}^{(3)}$, and the equations of motion for the meson fields are reduced to
\begin{eqnarray}
m^2_\sigma \overline{\sigma}&=&g_\sigma[ n^s-b_\sigma M(g_\sigma \overline{\sigma})^2-c_\sigma (g_\sigma \overline{\sigma})^3\notag\\
&&+C_{\delta\sigma}(g_\sigma \overline{\sigma})(g_\delta \overline{\delta}^{(3)})^2],\\
m^2_\omega \overline{\omega}_0&=&g_\omega[ n-c_\omega (g_\omega \overline{\omega}_0)^3-\Lambda_V (g_\omega \overline{\omega}_0)(g_\rho \overline{\rho}^{(3)}_0)^2],\\
m^2_\rho \overline{\rho}^{(3)}_0&=&g_\rho[ (n_p-n_n)-\Lambda_V (g_\rho \overline{\rho}^{(3)}_0)(g_\omega \overline{\omega}_0)^2],\\
m^2_\delta \overline{\delta}^{(3)}&=&g_\delta[ (n^s_p-n^s_n)+C_{\delta\sigma}(g_\delta \overline{\delta}^{(3)})(g_\sigma \overline{\sigma})^2],
\end{eqnarray}
where the nucleon scalar density $n^s$ and vector density are defined as
\begin{subequations}
	\begin{align}
	n^s=&\langle\overline{\Psi}\Psi\rangle=n^s_n+n^s_p, \\
	n=&\langle\overline{\Psi}\gamma^0\Psi\rangle=n_n+n_p
	\end{align}
\end{subequations}
with neutron scalar density $n^s_n$ and proton scalar density $n^s_p$ given by
\begin{eqnarray}
n^s_J&=&\frac{2}{(2\pi)^3}\displaystyle \int_{0}^{k^J_F}\frac{m^\ast_J}{\sqrt{(\vec k)^2+(m^\ast_J)^2}}\,d\vec k\notag\\
&=&\frac{m^\ast_J}{2\pi^2}[k^J_F\epsilon^{\ast,J}_F-(m^\ast_J)^2\ln\frac{k^J_F+\epsilon^{\ast,J}_F}{m^\ast_J}], J=n,p,
\end{eqnarray}
where $\epsilon^{\ast,J}_F$ and $k^J_F$ are expressed as
\begin{eqnarray}
\epsilon^{\ast,J}_F&=&\sqrt{(k^J_F)^2+(m^\ast_J)^2}, \notag\\
k^J_F&=&k_F(1+\tau^J_3\alpha)^\frac{1}{3},k_F=(\frac{3}{2}\pi^2 n)^\frac{1}{3}
\end{eqnarray}
with $k_F$ being the Fermi momentum of SNM. The effective (Dirac) masses of the neutron $m^\ast_n$ and proton $m^\ast_p$ can be expressed as
\begin{equation}
m^\ast_J=m-g_\sigma\overline{\sigma}-g_\delta\overline{\delta}^{(3)}\tau^J_3,J=n,p.
\end{equation}
In addition, the nucleon scalar self-energy and vector self-energy are given by
\begin{subequations}
	\begin{align}
	\Sigma_{s}^{J}=&-g_\sigma\overline{\sigma}-g_\delta\overline{\delta}^{(3)}\tau^J_3, \\
	\Sigma_{\mu}^{J}=&g_\omega\overline{\omega}_0+g_\rho\overline{\rho}^{(3)}_0\tau^J_3
	\end{align}
\end{subequations}
with $\tau_3 = 1 (-1)$ for protons (neutrons).

The set of coupled equations for the nucleon and meson
fields can be solved self-consistently using the iteration
method, and the properties of the nuclear matter can then
be obtained from these fields. According to the energy-momentum tensor
\begin{equation}
\mathcal{T}^{\mu\nu}=\sum_{\phi_a}\frac{\partial\mathcal{T}}{\partial\partial_\mu\phi_a}\partial^\nu\phi_a-\mathcal{T}g^{\mu\nu},
\end{equation}
the energy density $\epsilon$ and the pressure $P$ can be obtained as
\begin{eqnarray}
\epsilon&=&\langle\mathcal{T}^{00}\rangle= \epsilon^{\rm{kin}}_p+\epsilon^{\rm{kin}}_n\notag\\
&+&\frac{1}{2}m^2_\sigma(\overline\sigma)^2+\frac{1}{2}m^2_\omega(\overline\omega_0)^2+\frac{1}{2}m^2_\rho(\overline\rho^{(3)}_0)^2+\frac{1}{2}m^2_\delta(\overline\delta^{(3)})^2  \notag\\
&+&\frac{1}{3}mb_\sigma(g_\sigma\overline\sigma)^3+\frac{1}{4}c_\sigma(g_\sigma\overline\sigma)^4+\frac{3}{4}c_\omega(g_\omega\overline\omega_0)^4  \notag\\
&+&\frac{3}{2}\Lambda_V (g_\rho \overline{\rho}^{(3)}_0)^2(g_\omega \overline{\omega}_0)^2-\frac{1}{2}C_{\delta\sigma}(g_\delta \overline{\delta}^{(3)})^2(g_\sigma \overline{\sigma})^2
\end{eqnarray}
and
\begin{eqnarray}
P&=&\frac{1}{3}\sum_{j=1}^{3}\langle\mathcal{T}^{jj}\rangle =P^{\rm{kin}}_p+P^{\rm{kin}}_n\notag\\
&-&\frac{1}{2}m^2_\sigma(\overline\sigma)^2+\frac{1}{2}m^2_\omega(\overline\omega_0)^2+\frac{1}{2}m^2_\rho(\overline\rho^{(3)}_0)^2-\frac{1}{2}m^2_\delta(\overline\delta^{(3)})^2\notag\\
&-&\frac{1}{3}mb_\sigma(g_\sigma\overline\sigma)^3-\frac{1}{4}c_\sigma(g_\sigma\overline\sigma)^4+\frac{1}{4}c_\omega(g_\omega\overline\omega_0)^4\notag\\
&+&\frac{1}{2}\Lambda_V (g_\rho \overline{\rho}^{(3)}_0)^2(g_\omega \overline{\omega}_0)^2+\frac{1}{2}C_{\delta\sigma}(g_\delta \overline{\delta}^{(3)})^2(g_\sigma \overline{\sigma})^2,
\end{eqnarray}
where $\epsilon^{\rm{kin}}_J$ and $P^{\rm{kin}}_J$ are the kinetic
contributions to the energy densities and pressure of the nuclear matter, respectively, and they are given as follow
\begin{eqnarray}
\epsilon^{\rm{kin}}_J&=&\frac{2}{(2\pi)^3}\displaystyle \int_{0}^{k^J_F}\sqrt{(\vec k)^2+(m^\ast_J)^2}\,d\vec k\notag\\
&=&\frac{1}{(\pi)^2}\displaystyle \int_{0}^{k^J_F}k^2\sqrt{k^2+(m^\ast_J)^2}\,dk\notag\\
&=&\frac{1}{4}[3\epsilon^{\ast,J}_F n_J+m^\ast_J n^s_J]\,,J=n,p
\end{eqnarray}
and
\begin{eqnarray}
P^{\rm{kin}}_J&=&\frac{2}{3(2\pi)^3}\displaystyle \int_{0}^{k^J_F}\frac{(\vec k)^2}{\sqrt{(\vec k)^2+(m^\ast_J)^2}}\,d^3k\notag\\
&=&\frac{1}{4}[\epsilon^{\ast,J}_F n_J-m^\ast_J n^s_J]\,,J=n,p.
\end{eqnarray}

The binding energy per nucleon can be obtained from the
energy density via
\begin{equation}
E = E(n,\alpha=0)=\frac{\epsilon}{n}-m,
\end{equation}
and the analytical expressions of $E_{\rm sym}(n)$ and $L(n)$ can be obtained from their definitions Eqs.~(\ref{Eq:Esym_def}) and ~(\ref{Eq:L_def}) as
\begin{eqnarray}
\label{Eq:Esym}
E_{\rm{sym}}(n)&=&\frac{k^2_F}{6\epsilon^\ast_F}+\frac{1}{2}(\frac{g_\rho}{m^\ast_\rho})^2n\notag\\
&-&\frac{1}{2}(\frac{g_\delta}{m^\ast_\delta})^2\frac{(m^\ast_0)^2n}{(\epsilon^\ast_F)^2[1+(\frac{g_\delta}{m^\ast_\delta})^2A(k_F,m^\ast_0)]}
\end{eqnarray}
and
\begin{eqnarray}
\label{Eq:L}
L(n)&=&\frac{k^2_F[(\epsilon^\ast_F)^2+(m^\ast_0)^2]}{6(\epsilon^\ast_F)^3}+\frac{g^2_\sigma(m^\ast_0)^2k^2_Fn}{2Q_\sigma(\epsilon^\ast_F)^4}\notag\\
&+&\frac{3g^2_\rho n}{2Q_\rho}-\frac{3g^3_\omega g^4_\rho \Lambda_V\overline{\omega}_0n^2}{Q_\omega Q^2_\rho}\notag\\
&-&\frac{3g_\sigma g^2_\delta(m^\ast_0)^2n^2}{2(\epsilon^\ast_F)^3Q_\sigma Q_\delta}[\frac{2g_\sigma(m^\ast_0)^2}{(\epsilon^\ast_F)^2}+\frac{m^\ast_0\Gamma}{Q_\delta}-2g_\sigma]\notag\\
&-&\frac{3 g^2_\delta(m^\ast_0)^2n}{2(\epsilon^\ast_F)^2Q_\delta}[1-\frac{2k^2_F}{3(\epsilon^\ast_F)^2}-\frac{n\Phi}{2Q_\delta}]
\end{eqnarray}
where the $\Gamma$ and $\Phi$ are defined as
\begin{eqnarray}
\Gamma&=&3g_\sigma g^2_\delta[\frac{2n^s_0}{(m^\ast_0)^2}-\frac{3n}{m^\ast_0\epsilon^\ast_F}+\frac{m^\ast_0 n}{(\epsilon^\ast_F)^3}]+2C_{\delta\sigma}g^2_\delta g^2_\sigma \overline{\sigma}, \notag\\
\Phi&=&\frac{2g^2_\delta k^2_F}{(\epsilon^\ast_F)^3}
\end{eqnarray}
and
\begin{subequations}
	\begin{align}
	Q_\sigma=&m^2_\sigma+g^2_\sigma A(k_F,m^\ast_0)+2mb_\sigma g^3_\sigma \overline{\sigma}+3c_\sigma g^4_\sigma \overline{\sigma}^2 \\
	Q_\omega=&m^2_\omega+3c_\omega g^4_\omega \overline{\omega}_0^2 \\
	Q_\rho=&(m^\ast_\rho)^2=m^2_\rho+\Lambda_V g^2_\rho g^2_\omega \overline{\omega}_0^2 \\
	Q_\delta=&(m^\ast_\delta)^2+g^2_\delta A(k_F,m^\ast_0)
	\end{align}
\end{subequations}
In addition, the $(m^\ast_\delta)^2$ and $A(k_F,m^\ast_0)$ are given by
\begin{eqnarray}
(m^\ast_\delta)^2=m^2_\delta-C_{\delta\sigma}g^2_\delta g^2_\sigma\overline{\sigma}^2
\end{eqnarray}
and
\begin{eqnarray}
A(k_F,m^\ast_0)&=&\frac{4}{(2\pi)^3}\displaystyle \int_{0}^{k_F}\frac{(\vec k)^2}{\sqrt{[(\vec k)^2+(m^\ast_0)^2]^\frac{3}{2}}}\,d^3k\notag\\
&=&3(\frac{n^s}{m^\ast_0}-\frac{n}{\epsilon^\ast_F})
\end{eqnarray}
with $\epsilon^\ast_F=\sqrt{k^2_F+(m^\ast_0)^2}$ and $m^\ast_0=m-g_\sigma\overline{\sigma}$ is the nucleon Dirac mass in symmetric nuclear matter.

The analytical expressions of $E_{\rm{sym}}(n)$ (Eq.~(\ref{Eq:Esym})) and $L(n)$ (Eq.~(\ref{Eq:L})) are very useful and convenient for exploring the properties of asymmetric nuclear matter. We note that the expression Eq.~(\ref{Eq:L}) is an extension of the analytical expression for $L(n)$ in the nonlinear RMF model (e.g., the FSUGold model) in Refs.~\citep{2012PhLB..711..104C} by considering the contribution of $\delta$-$\sigma$ mixing.

For NSs, we adopt here the conventional NS model, i.e., the NS contains core, inner crust and outer crust, and the core is assumed to be composed of only neutrons, protons, electrons and possible muons~($npe\mu$) and its EOS can be calculated within the RMF models. For the details, one can refer to Refs.~\citep{2019PhRvD..99l1301Z,2019ApJ...886...52Z,2009ApJ...697.1549X,2016PhRvC..94f4326Z}. We would like to point out that in this work, the core-crust transition density $n_t$, which separates the liquid core from the solid inner crust of neutron stars, is obtained self-consistently by the thermodynamical method~\citep{2009ApJ...697.1549X,2007PhRvC..76b5801K,2012PhRvC..85b4302C}. In addition, the causality condition is satisfied for all the EOSs used in the following NS calculations.

\section{Results and Discussions}

\begin{table*}[t!]
	\centering
	\caption{Model parameters for the four models FSUGold, FSU-J0, FSU-$\delta6.7$ and FSU-$\delta6.2$.}
	\begin{tabular}{|c|c|c|c|c|c|c|c|c|c|}
		\hline
		Model           & $g_\sigma$ & $g_\omega$ & $g_\rho$ & $g_\delta$ & $b_\sigma(10^{-3})$ & $c_\sigma(10^{-3})$ & $c_\omega(10^{-2})$ & $\Lambda_V$ & $C_{\delta \sigma}(10^{-2})$ \\ \hline\hline
		FSUGold         & 10.5924    & 14.3020    & 5.88367  & -          & 0.756283            & 3.96033             & 1.00000             & 0.240000    & -                            \\ \hline
		FSU-J0       & 10.2143    & 13.4353    & 5.31345  & -          & 1.59524             & 0.540269            & 0.528340            & 0.171042    & -                            \\ \hline
		FSU-$\delta6.7$    & 10.2143    & 13.4353    & 7.26866  & 6.70000    & 1.59524             & 0.540269            & 0.528340            & 0.0214000   & 3.85000                      \\ \hline
		FSU-$\delta$6.2 & 10.2143    & 13.4353    & 6.98518  & 6.20000    & 1.59524             & 0.540269            & 0.528340            & 0.0200112   & 5.40000                      \\ \hline
	\end{tabular}
	\label{TABLE_parameter}
\end{table*}

\begin{table}[t!]
	\centering
	\caption{Nuclear matter bulk parameters, NS properties and ground state properties of $^{208}$Pb and $^{48}$Ca in FSUGold, FSU-J0, FSU-$\delta6.7$ and FSU-$\delta6.2$. For comparison, the nuclear ground state data are listed as following: the binding energy $E^{208}_B = -7.87$~MeV and $E^{48}_B = -8.67$~MeV~\citep{Wang:2021xhn}, the charge rms radius $R^{208}_{\rm ch} = 5.50$~fm and $R^{48}_{\rm ch} = 3.48$~fm~\citep{Angeli:2013epw}, and the spin-orbit splitting $\epsilon^{208}_{ls,2d\pi} = 1.32$~MeV, $\epsilon^{208}_{ls,3p\nu} = 0.89$~MeV and $\epsilon^{208}_{ls,2f\nu} = 1.77$~MeV~\citep{1972PhRvC...5..626V}.}
	\begin{tabular}{|c|c|c|c|c|}
		\hline
		Quantity                          & FSUGold & FSU-J0 & FSU-$\delta6.7$ & FSU-$\delta$6.2 \\ \hline
		$n_0(\rm{fm}^{-3})$               & 0.148   & 0.148  & 0.148           & 0.148           \\ \hline
		$e_0(\rm{MeV})$                   & -16.31  & -16.31 & -16.31          & -16.31          \\ \hline
		$K_0(\rm{MeV})$                   & 229.2   & 229.2  & 229.2           & 229.2           \\ \hline
		$J_0(~\rm{MeV})$                  & -521.6  & -322.0 & -322.0          & -322.0          \\ \hline
		$m^\ast_{\rm{Dirac}}$             & 0.610   & 0.610  & 0.610           & 0.610           \\ \hline
	    $m^\ast_{\rm{Dirac},n,0.5}$           & 0.610   & 0.610  & 0.576           & 0.578           \\ \hline
		$m^\ast_{\rm{Dirac},p,0.5}$           & 0.610   & 0.610  & 0.643           & 0.640           \\ \hline
		$E_{{\rm{sym}},c}(\rm{MeV})$      & 25.88   & 25.88  & 25.88           & 25.88           \\ \hline
		$L_c(\rm{MeV})$                   & 48.02   & 55.00  & 55.00           & 55.00           \\ \hline
		$E_{{\rm{sym}}}(\rm{MeV})$        & 32.59   & 33.45  & 32.75           & 32.53           \\ \hline
		$L(\rm{MeV})$                     & 60.50   & 68.14  & 53.50           & 48.21           \\ \hline
		$n_t(\rm{fm}^{-3})$               & 0.082   & 0.081  & 0.088           & 0.089           \\ \hline
		$n_{\rm{cen}}(\rm{fm}^{-3})$      & 1.15    & 1.00   & 0.91            & 0.92            \\ \hline
		$R_{1.4}(\rm{km})$                & 12.66   & 13.65  & 13.67           & 12.76           \\ \hline
		$\Lambda_{1.4}$                   & 401.0   & 576.0  & 578.5           & 420.0           \\ \hline
		$M_{\rm{max}}$($M_{\odot}$)       & 1.72    & 1.93   & 2.05            & 2.10            \\ \hline
		$E^{208}_B(\rm{MeV})$                   & -7.89   & -7.94  & -7.69           & -7.71           \\ \hline
		$R^{208}_{\rm ch}(\rm{fm})$                 & 5.54    & 5.55   & 5.55            & 5.54            \\ \hline
		$\Delta r^{208}_{\rm{np}}(\rm{fm})$     & 0.207   & 0.227  & 0.235           & 0.242           \\ \hline
		$\epsilon^{208}_{ls,2d\pi}(\rm{MeV})$ & 1.60    & 1.57   & 1.55            & 1.55            \\ \hline
		$\epsilon^{208}_{ls,3p\nu}(\rm{MeV})$ & 0.77    & 0.75   & 0.71            & 0.71            \\ \hline
		$\epsilon^{208}_{ls,2f\nu}(\rm{MeV})$ & 2.06    & 1.98   & 1.96            & 1.95            \\ \hline
		$E^{48}_B(\rm{MeV})$                   & -8.56   & -8.61  & -8.46           & -8.47           \\ \hline
		$R^{48}_{\rm ch}(\rm{fm})$                 & 3.46    & 3.49   & 3.49            & 3.49            \\ \hline
		$\Delta r^{48}_{\rm{np}}(\rm{fm})$     & 0.197   & 0.208  & 0.213           & 0.217           \\ \hline
	\end{tabular}
    \label{TABLE_properties}
\end{table}

The FSUGold interaction~\citep{2005PhRvL..95l2501T} is an accurately calibrated relativistic parametrization in the non-linear RMF model and it can very successfully describe the ground state binding energy and charge radius of finite nuclei as well as their linear response.
In the FSUGold model~\citep{2005PhRvL..95l2501T,2010PhRvC..82e5803F},
the nucleon mass is fixed at $m=939.0~\rm{MeV}$, and the three meson masses are also fixed, namely, $m_\sigma=491.5~\rm{MeV}$, $m_\omega=782.5~\rm{MeV}$ and $m_\rho=763.0~\rm{MeV}$. In this case, the FSUGold has thus seven model parameters, i.e.,
$g_\sigma$, $g_\omega$, $g_\rho$, $b_\sigma$, $c_\sigma$, $c_\omega$ and $\Lambda_V$, which are listed in TABLE~\ref{TABLE_parameter}.
The bulk properties of infinite nuclear matter of FSUGold can be obtained as: $n_0=0.148~\rm{fm}^{-3}$, the binding energy per nucleon of SNM at $n_0$ is $e_0 \equiv E_0(n_0) = -16.31~\rm{MeV}$, $K_0=229.2~\rm{MeV}$, the nucleon Dirac mass in SNM at $n_0$ is $m^\ast_{\rm{Dirac}}=m^\ast_0/m=0.61$, the neutron (proton) Dirac mass in asymmetric nuclear matter at $n_0$ and $\alpha=0.5$ is $m^\ast_{\rm{Dirac,n,0.5}}=m^\ast_n(n=n_0,\alpha=0.5)/m=0.61$ ($m^\ast_{\rm{Dirac,p,0.5}}=m^\ast_p(n=n_0,\alpha=0.5)/m=0.61$),
$J_0=-521.6~\rm{MeV}$, $E_{{\rm{sym}},c}\equiv E_{\rm{sym}}(n_c) =25.88~\rm{MeV}$ and $L_c=48.02~\rm{MeV}$.
In addition, the FSUGold predicts $E_{\rm{sym}}(n_0)=32.59$~MeV, $L=60.5$~MeV, $n_t=0.082~\rm{fm}^{-3}$, the central density of the maximum mass NS configuration $n_{\rm cen}=1.15~\rm{fm}^{-3}$, $R_{1.4} = 12.66$~km, $\Lambda_{1.4} = 401$, and $M_{\rm max} = 1.72M_{\odot}$.
TABLE~\ref{TABLE_properties} summarizes these results.
Also included in TABLE~\ref{TABLE_properties} are the FSUGold RMF predictions of some ground state properties of doubly magic nuclei $^{208}$Pb and $^{48}$Ca, i.e., the binding energy per nucleon $E^{A}_B$, the charge rms radius $R^{A}_{\rm ch}$, the neutron skin thickness $\Delta r^{A}_{\rm np}$ with $A = 208~(48)$ for $^{208}$Pb~($^{48}$Ca), as well as the $^{208}$Pb spin-orbit splittings of proton $2d$ state $\epsilon^{208}_{ls,2d\pi}$, neutron $3p$ state $\epsilon^{208}_{ls,3p\nu}$ and neutron $2f$ state $\epsilon^{208}_{ls,2f\nu}$.

One sees from TABLE~\ref{TABLE_properties} that while the FSUGold nicely reproduces the ground state properties of finite nuclei, it predicts a too small value of $M_{\rm max} = 1.72M_{\odot}$, which is significantly smaller than the observed $2.08^{+0.07}_{-0.07}M_{\odot}$ for PSR J0740+6620~\citep{2020NatAs...4...72C,2021ApJ...915L..12F}.
In addition, we note the
$\Delta r_{\rm{np}}$ of $^{208}$Pb is $0.207$~fm with the FSUGold, which is smaller than the PREX-II measurement.
Noting the variation of
the higher-order bulk parameters (e.g., $J_0$) essentially does not change the properties of finite nuclei~\citep{2019PhRvD..99l1301Z}, we thus increase $J_0$ while keep $n_0$, $e_0$, $K_0$, $m^\ast_{\rm{Dirac}}$ and $E_{\rm{sym},c}$ unchanged to enhance the $M_{\rm max}$. At the same time, we increase $L_c$ from original $48.02~\rm{MeV}$ to $55~\rm{MeV}$ to fit the constraint of $L_c \gtrsim 52~\rm{MeV}$~\citep{2021arXiv210205267Y} from PREX-II. The $L_c = 55~\rm{MeV}$ corresponds to $L(2/3n_0) = 54~\rm{MeV}$, again consistent with the constraint of $L(2/3n_0) \gtrsim 49$~MeV~\citep{2021PhRvL.126q2503R} from PREX-II.
Indeed, the $L_c=55~\rm{MeV}$ leads to $\Delta r_{\rm{np}} \approx 0.22$~fm for $^{208}$Pb according to the relation between $\Delta r_{\rm{np}}$ of $^{208}$Pb and $L_c$, i.e., $\Delta r_{\rm{np}} = (0.0615\pm 0.0022) + (0.00282\pm 0.0000344)L_c$~\citep{2021arXiv210205267Y}. In addition, we note that the $L_c=55~\rm{MeV}$ produces $\Lambda_{1.4} = 576$, and $L_c > 56~\rm{MeV}$ will violate the constraint $\Lambda_{1.4}\leq580$.
Increasing $J_0$ can lead to larger $M_{\rm max}$, but the $J_0$ cannot be larger than $-322.0~\rm{MeV}$ to fit the constraint on the SNM EOS from flow data~\citep{2002Sci...298.1592D}. One then obtains a value of $M_{\rm max} = 1.93M_{\odot}$ with $J_0 = -322.0~\rm{MeV}$, suggesting the FSUGold Lagrangian cannot simultaneously describe the observed $2.08^{+0.07}_{-0.07}M_{\odot}$ for PSR J0740+6620~\citep{2020NatAs...4...72C,2021ApJ...915L..12F} and the constraint on the SNM EOS from flow data~\citep{2002Sci...298.1592D}.
From the seven bulk parameters $n_0=0.148~\rm{fm}^{-3}$, $e_0 = -16.31~\rm{MeV}$, $K_0=229.2~\rm{MeV}$, $m^\ast_{\rm{Dirac}} = 0.61$, $J_0 = -322.0~\rm{MeV}$, $E_{\rm{sym},c} =25.88~\rm{MeV}$ and $L_c=55~\rm{MeV}$, one can obtain the seven model parameters $g_\sigma$, $g_\omega$, $g_\rho$, $b_\sigma$, $c_\sigma$, $c_\omega$ and $\Lambda_V$, and the corresponding parameter set is denoted as `FSU-J0' and shown in TABLE~\ref{TABLE_parameter}. Moreover, the nuclear matter bulk parameters and NS properties as well as the ground state properties of $^{208}$Pb and $^{48}$Ca with FSU-J0 are summarized in TABLE~\ref{TABLE_properties}.
As expected, one sees that compared to FSUGold, the FSU-J0 gives very similar ground state properties except a larger neutron skin thickness of $\Delta r^{208}_{\rm np} = 0.227$~fm due to the larger $L_c$ and thus it is compatible with the PREX-II measurement.

Now we turn to the Lagrangian density $\mathcal{L}$ given by Eq.~(\ref{Eq:Lagrangian}), which additionally includes the $\delta$ meson and the $\delta$-$\sigma$ mixing compared with the FSUGold model and has nine model parameters, namely $g_\sigma$, $g_\omega$, $g_\rho$, $g_\delta$, $b_\sigma$, $c_\sigma$, $c_\omega$, $\Lambda_V$ and $C_{\delta\sigma}$ with the $\delta$ meson mass fixed at $980.0~\rm{MeV}$.
To explore the effects of the $\delta$ meson and the $\delta$-$\sigma$ mixing, our strategy is to vary the $g_\delta$ and $C_{\delta\sigma}$ while fix the values of $n_0$, $e_0$, $K_0$, $J_0$, $m^\ast_{\rm{Dirac}}$, $E_{\rm{sym},c}$ and $L_c$ at their corresponding values as in FSU-J0, and this guarantees that the constraints from PREX-II and SNM EOS from flow data can be satisfied as well as the basic properties of finite nuclei can be reasonably described.
In such a way,
we find that increasing the value of $g_\delta$ can stiffen the symmetry energy at suprasaturation densities and enhance the $M_{\rm max}$. However, the $g_\delta$ cannot be too large since the $\Lambda_{1.4}$ will also increase with $g_\delta$ but it is limited by $\Lambda_{1.4} \le 580$.
On the other hand, it is interesting to note that the $C_{\delta\sigma}$ can soften the symmetry energy at intermediate densities around $2\sim 3n_0$ but stiffen the symmetry energy at high densities, and this feature can decrease the $\Lambda_{1.4}$ and increase the $M_{\rm max}$. As an example, we find the parameters $g_\delta=6.7$ and $C_{\delta\sigma}=0.0385$ lead to $\Lambda_{1.4} = 578.5$ and $M_{\rm max} = 2.05M_{\odot}$, and we denote this parameter set as `FSU-$\delta6.7$' and the model parameters are listed in TABLE~\ref{TABLE_parameter}. The nuclear matter bulk parameters and NS properties as well as the ground state properties of $^{208}$Pb and $^{48}$Ca with FSU-$\delta6.7$ are summarized in TABLE~\ref{TABLE_properties}.
One sees that the FSU-$\delta6.7$ predicts an isospin splitting of the nucleon Dirac mass with
$m^\ast_{\rm{Dirac},p,0.5} > m^\ast_{\rm{Dirac},n,0.5}$ due to the inclusion of the $\delta$ meson, which is not observed in the FSUGold and FSU-J0.
As for the ground state properties, the FSU-$\delta6.7$ reproduces the experimental data on binding energies at the $2\%$ level and charge rms radii better than $1\%$, which is already satisfactory in the present work without the best-fit procedure.
In addition, it is interesting to see that the inclusion of the $\delta$ meson seems to slightly improve the description of the spin-orbit splittings and obviously enhance the neutron skin thickness.

\begin{figure}[t!]
	\centering
	\includegraphics[width=0.85\linewidth]{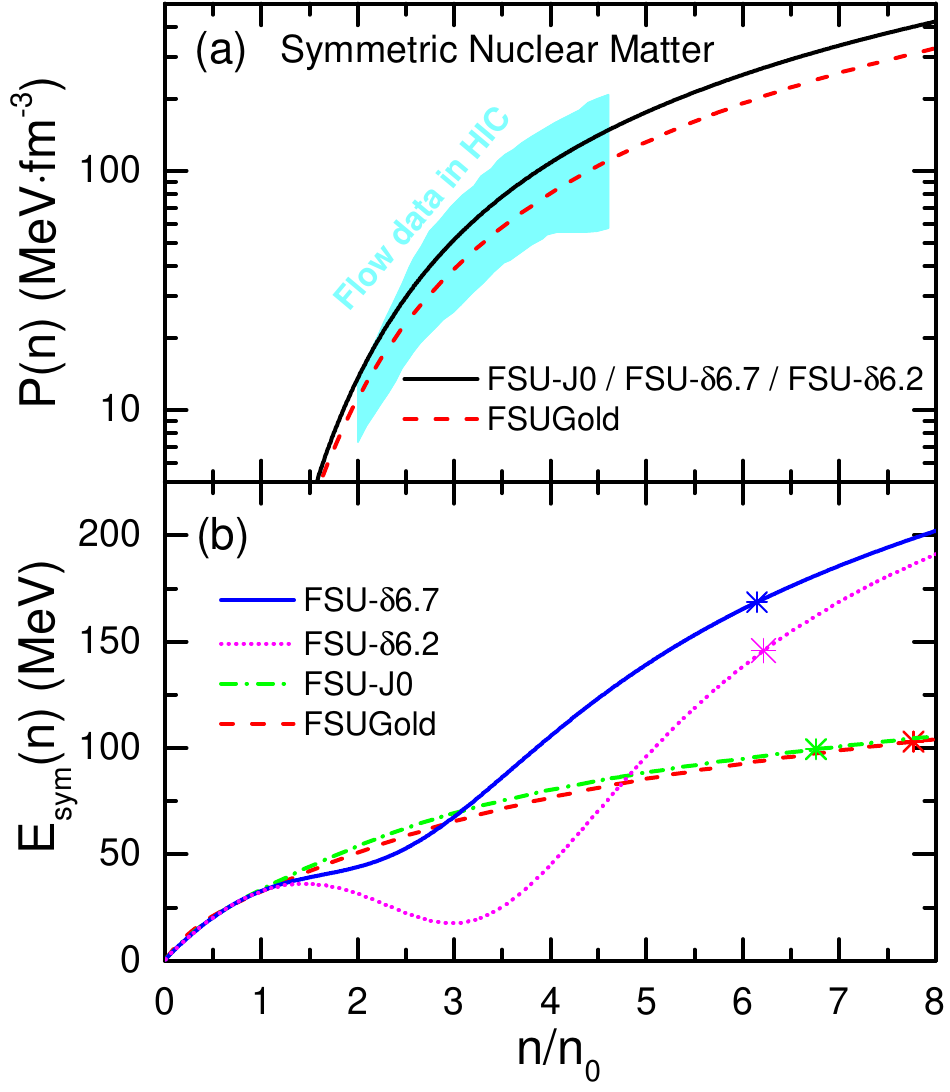}
	\caption{Pressure vs density (a) and density dependence of the symmetry energy (b) predicted by FSUGold, FSU-J0, FSU-$\delta6.7$ and FSU-$\delta6.2$. The shaded area in panel (a) represents the constraints from flow data in heavy-ion collisions~\citep{2002Sci...298.1592D}.
		The central density of maximum mass NS configuration is indicated by `$\ast$' in panel (b).}
	\label{Fig_EOS}
\end{figure}

As a more interesting example,
we find the parameters $g_\delta=6.2$ and $C_{\delta\sigma}=0.054$ lead to $\Lambda_{1.4} = 420$ and $M_{\rm max} = 2.1M_{\odot}$. This parameter set is denoted as `FSU-$\delta6.2$' and its model parameters are listed in TABLE~\ref{TABLE_parameter}. The nuclear matter bulk parameters and NS properties as well as the ground state properties of $^{208}$Pb and $^{48}$Ca with FSU-$\delta6.2$ are also included in TABLE~\ref{TABLE_properties} for comparison.
It is seen that the FSU-$\delta6.2$ predicts almost the same isospin splitting of nucleon Dirac mass and nuclear ground state properties as those with the FSU-$\delta6.7$ except a larger neutron skin thickness.
The important feature of FSU-$\delta6.2$
is that it gives a very soft symmetry energy around $2\sim 3n_0$ due to the smaller $g_\delta$ and larger $C_{\delta\sigma}$ compared with FSU-$\delta6.7$. It should be mentioned that a larger $C_{\delta\sigma}$ value may lead to extremely soft symmetry energy around intermediate densities and cause the spinodal instability for the NS matter, i.e., the pressure decreases
with the increment of density, and this case needs special treatment with the construction of two-phase system in NSs~\citep{2019PhRvC..99c5209Z,2020PhRvC.102f5803K}.

To illustrate the above discussions and analyses,
we show in Fig.~\ref{Fig_EOS}~(a) the pressure vs density for SNM and (b) the $E_{\rm{sym}}$ with FSUGold, FSU-J0, FSU-$\delta6.7$ and FSU-$\delta6.2$. The constraint on pressure vs density from collective flow data in heavy-ion collisions~\citep{2002Sci...298.1592D} is also included in Fig.~\ref{Fig_EOS}~(a) for comparison.
Indeed, one sees from Fig.~\ref{Fig_EOS}~(a) that the results from all the four parameter sets FSUGold, FSU-J0, FSU-$\delta6.7$ and FSU-$\delta6.2$ are consistent with the constraint on the SNM EOS from collective flow data.
In addition,
it is seen from Fig.~\ref{Fig_EOS}~(b) that compared with FSU-J0, the FSU-$\delta6.7$ predicts a softer symmetry energy around $2n_0$ but much stiffer symmetry energy at high densities due to the effects of $\delta$ meson and the $\delta$-$\sigma$ mixing, which leads to a larger $M_{\rm max}$ for FSU-$\delta6.7$ although both FSU-J0 and the FSU-$\delta6.7$ have the same EOS of SNM.
Furthermore, the FSU-$\delta6.2$ predicts a very soft symmetry energy around $2\sim 3n_0$ but a very stiff symmetry energy at high densities where the symmetry energy increases rapidly with density (resulting in a larger sound speed in the NS matter), and this produces an even larger $M_{\rm max}$ (i.e., $2.1M_{\odot}$) compared with FSU-$\delta6.7$. The very soft symmetry energy around $2\sim 3n_0$ with FSU-$\delta6.2$ produces a much smaller $\Lambda_{1.4}$ (i.e., $420$).
These results clearly indicate that the $\delta$ meson and the $\delta$-$\sigma$ mixing can have very interesting and significant influence on the symmetry energy and the NS properties.

\begin{figure}[t!]
	\centering
	\includegraphics[width=0.9\linewidth]{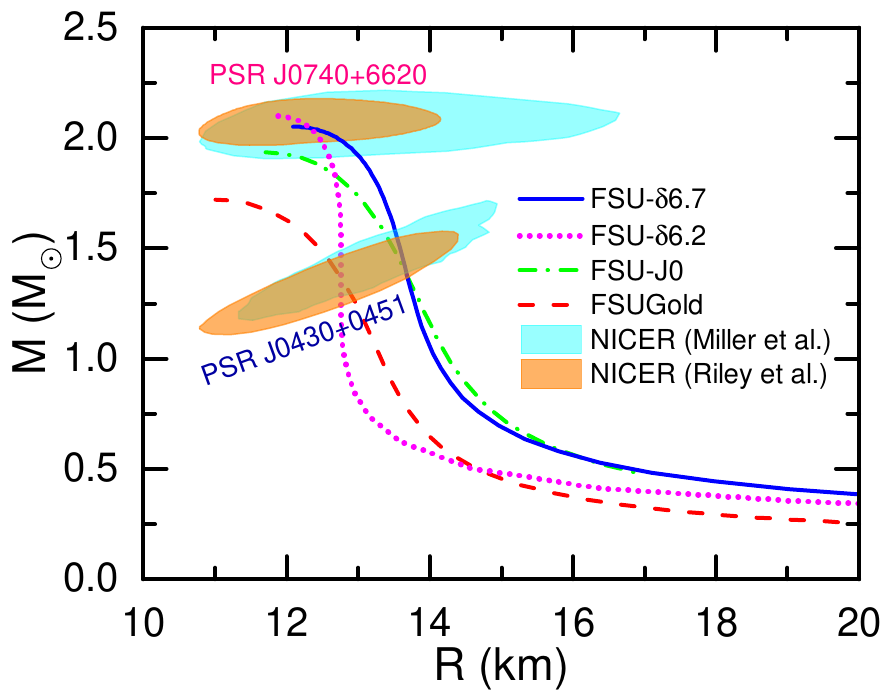}
	\caption{M-R relation for static NSs from FSUGold, FSU-J0, FSU-$\delta6.7$ and FSU-$\delta6.2$. The NICER (XMM-Newton) constraints with $68\%$~C.L for PSR J0030+0451~\citep{2019ApJ...887L..21R,2019ApJ...887L..24M} and PSR J0740+6620~\citep{2021ApJ...918L..27R,2021ApJ...918L..28M} are also included for comparison.}
	\label{Fig_MR}
\end{figure}

Finally, it is interesting to check the model predictions on the NS M-R relation
since the simultaneous M-R determinations from NICER for PSR J0030+0451~\citep{2019ApJ...887L..21R,2019ApJ...887L..24M} with mass around $1.4M_{\odot}$ and for PSR J0740+6620~\citep{2021ApJ...918L..27R,2021ApJ...918L..28M} with mass around $2.0M_{\odot}$ have been obtained.
Shown in Fig.~\ref{Fig_MR}
is the M-R relation for static NSs predicted from the four models FSUGold, FSU-J0, FSU-$\delta6.7$ and FSU-$\delta6.2$.
Also included in Fig.~\ref{Fig_MR} are
the two independent simultaneous M-R measurements from NICER by analyzing the X-ray data for the isolated $205.53$~Hz millisecond pulsar PSR J0030+0451~\citep{2019ApJ...887L..21R,2019ApJ...887L..24M}
as well as the recent radius determination of PSR J0740+6620 from NICER and X-ray Multi-Mirror (XMM-Newton) X-ray
observations~\citep{2021ApJ...918L..27R,2021ApJ...918L..28M}.
It is seen that the FSU-$\delta6.7$ and FSU-$\delta6.2$, which consider the $\delta$ meson and the $\delta$-$\sigma$ mixing, are in very good agreement with the astrophysical observations and measurements for both PSR J0030+0451 and PSR J0740+6620.
On the other hand, the FSUGold and FSU-J0 within the original FSUGold framework fail to describe the constraints for PSR J0740+6620 due to the too small value of $M_{\rm max}$.
At last,
we would like to point out that the present
work is based on the conventional NS model within a single
unified framework without considering the possible appearance of new
degrees of freedom such as hyperons, meson condensates, quark
matter in NSs. It will be interesting to consider these new
degrees of freedom in NSs within the new RMF model proposed in the present work.

\section{Summary}
By including the isovector scalar $\delta$ meson and its coupling to isoscalar scalar $\sigma$ meson in the accurately calibrated interaction FSUGold, we have demonstrated that the new RMF model can
be simultaneously compatible with
the constraints on the EOS of SNM at suprasaturation densities from flow data in heavy-ion collisions,
the neutron skin thickness of $^{208}$Pb from the PREX-II experiment,
the measured largest NS mass reported so far from PSR J0740+6620,
the limit of $\Lambda_{1.4}\leq580$ from GW170817,
and the M-R relation of PSR J0030+0451 and PSR J0740+6620 measured by NICER,
and thus remove the tension between PREX-II and GW170817 observed in the conventional RMF model.

The great success of the new RMF model is mainly due to the fact that
the $\delta$-$\sigma$ mixing can soften the symmetry energy at intermediate densities but stiffen the symmetry energy at high densities,
which further leads to a larger $M_{\rm max}$ but a smaller $\Lambda_{1.4}$.
Further investigation of the $\delta$-$\sigma$ mixing effects on nuclear structure,
heavy-ion collisions and neutron stars is an extremely interesting topic.

\section*{Acknowledgements}
This work was supported by National SKA Program of China No. 2020SKA0120300 and the National Natural Science Foundation of China under Grant No. 11625521 and No. 11775049.


\bibliography{NSDelta}{}
\bibliographystyle{aasjournal}



\end{document}